\documentclass[a4paper,twocolumn,showpacs,prl]{revtex4}%
\usepackage{amsfonts}
\usepackage{amsmath}
\usepackage{amssymb}
\usepackage{graphicx}
\usepackage{amsbsy}%
\providecommand{\U}[1]{\protect \rule{.1in}{.1in}}

\begin{document}
\title{Quantum tunneling through planar p-n junctions in HgTe quantum wells}
\author{L. B. Zhang}
\affiliation{SKLSM, Institute of Semiconductors, Chinese Academy of Sciences, P.O. Box 912,
Beijing 100083, China}
\author{Kai Chang}
\email{kchang@red.semi.ac.cn}
\affiliation{SKLSM, Institute of Semiconductors, Chinese Academy of Sciences, P.O. Box 912,
Beijing 100083, China}
\author{X. C. Xie}
\affiliation{Department of Physics, Oklahoma State University, Stillwater, Oklahoma 74078, USA}
\author{H. Buhmann, and L. W. Molenkamp}
\affiliation{Physikalisches Institut (EP3), Universitat Wuerzburg, Am Hubland, D-97074
Wuerzburg, Germany}
\date{\today }


\begin{abstract}
We demonstrate that a p-n junction created electrically in HgTe quantum wells
with inverted band-structure exhibits interesting intraband and interband
tunneling processes. We find a perfect intraband transmission for electrons
injected perpendicularly to the interface of the p-n junction. The opacity and
transparency of electrons through the p-n junction can be tuned by changing
the incidence angle, the Fermi energy and the strength of the Rashba
spin-orbit interaction. The occurrence of a conductance plateau due to the
formation of topological edge states in a quasi-one-dimensional p-n junction
can be switched on and off by tuning the gate voltage. The spin orientation
can be substantially rotated when the samples exhibit a moderately strong
Rashba spin-orbit interaction.

\end{abstract}
\maketitle

Electrical manipulation of the transport property of topological
insulators is becoming one of the central issues in condensed matter
physics and has attracted rapid growing interests. Instead of
symmetry breaking and a local order parameter, topological
insulators are characterized by a topological
invariant.\cite{QH1982,Haldane,Niu} Originally, graphene was
proposed as a quantum spin Hall (QSH) insulator,\cite{Kane-Mele} but
this turned out to be difficult to realize because the spin-orbit
interaction (SOI) in this material is too small to create a gap
observable in transport experiments. The QSH insulator state was
proposed independently to occur in HgTe quantum wells (QWs) with
inverted band structure \cite{BHZ,Kai} and demonstrated in a recent
experiment.\cite{Marcus} This nontrivial topological
insulator\cite{LK,C2,C3,C4,C5,Kane,Bernevig} can be distinguished
from an ordinary band insulator (BI) by the existence of a $Z_{2}$
topological invariant.\cite{Kane,LK} It possesses an insulating bulk
and metallic edges, where electrons with up and down spins
counter-propagate along opposite edges, and consequently shows a
quantized conductance when the Fermi energy is swept across the bulk
gap.\cite{Marcus} These topological edge states are robust against
local perturbations, e.g., impurity scattering\cite{Kane,Shen,Xie}
processes and the Coulomb interaction\cite{Xu-Moore,B-Z}, and lead
to novel transport properties. The low energy spectrum of carriers
in QWs with inverted band structure, e.g., a TI, can be well
described by a four band Hamiltonian.\cite{BHZ} It is highly
desirable to study how to manipulate the transport properties
electrically, both from a basic physics and a device application
perspective.

In this Letter we investigate quantum tunneling through planar p-n
junctions in HgTe QWs with inverted band structure. Perfect
tunneling transmission is found for electrons incident at normal
angles. The opacity and transparency of the tunnel barrier can be
controlled by tuning the angle of incidence of the charge carriers,
the gate voltage (Fermi energy) and the SOI. An interesting spin
refraction effect is found utilizing the strong Rashba spin-orbit
interaction (RSOI) in HgTe QWs, i.e., the tunnel barrier is
transparent for one spin orientation and opaque for the other. This
phenomenon is difficult to be observed in a conventional
semiconductor two-dimensional electron gas due to the weak
SOI.\cite{spinoptics} The quantum states in HgTe QWs consist of
electron and heavy hole states and therefore show a very strong SOI
which is one and even two orders magnitude than that in conventional
semiconductor 2DEG, and therefore makes it possible to observe spin
refraction. The topological edge channels lead to the plateau of the
quantized conductance in the p-n junction sample with narrow
transverse width, but the spin orientation can be rotated
significantly by the RSOI. The conductance plateau can be destroy
when the QSH topological insulator is driven into the normal
insulator by tuning the external electric field.

A planar p-n junction, as schematically shown in Fig. 1, can be
fabricated using top gates inducing an electrostatic potential in
the electron gas underneath. Electrons are injected from a nearby
quantum point contact (QPC), and transmitted and/or reflected at the
interface of the p-n junction. The transmitted and/or reflected
electrons can be collected by other QPCs to provide further angular
information. The angle of incidence angle of the electrons can be
tuned by applying a perpendicular magnetic field.

The low energy spectrum of carriers in a HgTe QW with inverted band
structure, including the RSOI, is
\begin{align*}
H  &  =\left(
\begin{array}
[c]{cc}%
H(k) & H_{RSOI}(k)\\
H_{RSOI}^{\ast}(-k) & H^{\ast}(-k)
\end{array}
\right) \\
&  =\left(
\begin{array}
[c]{cccc}%
\epsilon_{k}+M(k) & Ak_{-} & i\alpha k_{-} & 0\\
Ak_{+} & \epsilon_{k}-M(k) & 0 & 0\\
-i\alpha k_{+} & 0 & \epsilon_{k}+M(k) & -Ak_{+}\\
0 & 0 & -Ak_{-} & \epsilon_{k}-M(k)
\end{array}
\right)
\end{align*}
where $\  \mathbf{k}=(k_{x},k_{y})$ is the carriers' in-plane momentum,
$\epsilon_{(}k)=C-D(k_{x}^{2}+k_{y}^{2})$, $M(k)=M-B(k_{x}^{2}+k_{y}^{2})$,
$k_{\pm}=k_{x}\pm ik_{y}$, and $A$, $B$, $C$, $D$, $M$ are the parameters
describing the band structure of the HgTe QWs. Note that the insulator state
characterized by the parameter $M$, which is determined by the thickness of
HgTe QW \cite{BHZ} (for negative (positive) $M$, the QW is a TI (BI),
respectively) Furthermore, $H(k)=\epsilon_{k}I_{2\times2}+d_{i}(k)\sigma_{i}$,
with $d_{1}=Ak_{x}$, $d_{2}=Ak_{y}$, $d_{3}=M(k)$. Without RSOI, i.e.,
$\alpha=0$, the eigenvalues and eigenvectors are $\epsilon_{\pm}=\epsilon
_{k}\pm \sqrt{M(k)^{2}+A^{2}k^{2}}$, $\chi_{\pm}=N_{\pm}(Ake^{-i\theta},\pm
d(k)\mp M(k))^{T}$, respectively, where $N_{\pm}=(A^{2}k^{2}+(d(k)\mp
M(k))^{2})^{-1/2}$ are the normalization constants, and $\theta$ is the
incidence angle. When the RSOI is included, i.e., $\alpha \neq0$, the
eigenvalues become $\epsilon_{+}^{\sigma}=\epsilon_{k}\pm \frac{k\alpha}{2}+
\sqrt{[M(k)\pm \frac{k\alpha}{2}]^{2}+A^{2}k^{2}}$, $\epsilon_{-}^{\sigma
}=\epsilon_{k}\pm \frac{k\alpha}{2}- \sqrt{[M(k)\pm \frac{k\alpha}{2}]^{2}%
+A^{2}k^{2}}$, respectively. The corresponding analytical expression of the
eigenvectors with RSOI are omitted here for simplicity.

We first model the electrostatic potential of a p-n junction by a step-like
potential, i.e., $V(x)=0$, for $x<0$; $V$ for $x\geqslant0$. This idealized
model can give the essential features of the quantum tunneling, a more
realistic smooth potential will be used in calculations aiming for comparison
with experiments (see Fig. 2). We assume electrons are injected from the QPC
at the left side of the junction with wave vector $k_{n}^{L}=k_{F}$, the wave
function at the left side is then $\psi_{L}(x<0)=[\chi_{n}^{L}e^{ik_{n}%
^{L}x\cos(\theta_{n}^{L})}+\sum_{m=1}^{4}r_{m}\chi_{m}^{L}e^{-ik_{m}^{L}%
x\cos(\theta_{m}^{L})}]e^{ik_{n}^{L}y\sin \theta_{n}^{L}}$, and at the right
side $\psi_{R}(x>0)=\sum_{m=1}^{4}t_{m}^{R}e^{ik_{m}^{R}x\cos(\theta_{m}^{R}%
)}e^{ik_{n}^{L}y\sin \theta_{n}^{L}}$, where $\chi_{n}^{\lambda}$($\lambda
=L,R$) is a four-component vector. The transmission $T(\theta)$ is obtained
using scattering matrix theory by matching the wave functions and the currents
at the interface of the p-n junction with conserved $p_{y}=k_{F}\sin \theta
_{n}^{L}$, i.e., $\psi_{L}(x<0)=\psi_{R}(x>0)$ and $j_{L}(x)\psi
_{L}(x<0)=j_{R}(x)\psi_{R}(x>0)$, where $j_{L,R}(x)$ are the current operators
along the propagation direction, i.e., the $x$-axis,
\begin{equation}
j_{L,R}(x)=\left(
\begin{array}
[c]{cccc}%
-2D_{+}k_{x} & A & i\alpha & 0\\
A & -2D_{-}k_{x} & 0 & 0\\
-i\alpha & 0 & -2D_{+}k_{x} & -A\\
0 & 0 & -A & -2D_{-}k_{x}%
\end{array}
\right)  ,\nonumber
\end{equation}
where $D_{+}=D+B$, $D_{-}=D-B$. Lengthy analytic expressions for the
transmission $T(\theta)$ and reflection coefficients can be obtained, but are
omitted here for brevity. The conductance can be calculated by $G=\int
_{-\pi/2}^{\pi/2}T(\theta)f(E_{F})d\theta$.

\begin{figure}[h]
\includegraphics[width=0.99\columnwidth,clip]{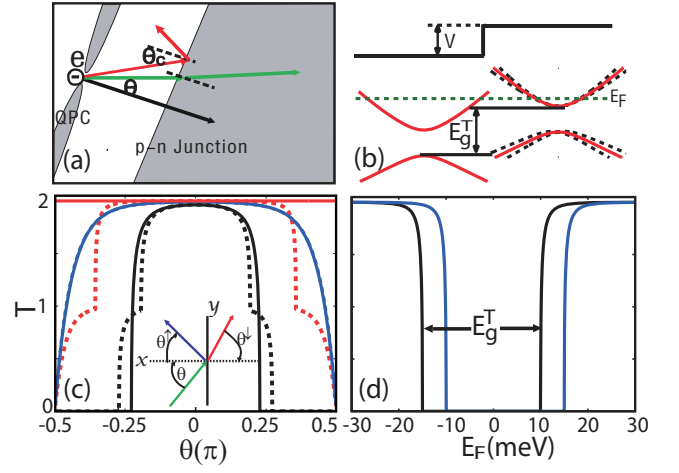} \caption{(color
online). (a) Schematic structure of a planar p-n junction in a HgTe QW with
inverted band structure. The electric gate is positioned above the shaded
region. (b) Schematic of the potential profile of the n-p junction and the
band structure of HgTe QWs without RSOI (red solid lines) and with RSOI (black
dashed lines). The green dashed line indicates the Fermi energy. $E_{g}^{T}$
indicates the gap in the transmission. (c) The transmission as a function of
incident angle $\theta$ with RSOI, $\alpha=50$ meV$\cdot$nm, (the dashed
curves) and without RSOI, $\alpha=0$, (the solid curves) for a given Fermi
energy $E_{F}=20$ meV. $V_{g}=5$, 0, -5 mV correspond to the blue, red and
black lines, respectively. The inset shows the spin refraction schematically.
(d) Transmission as a function of Fermi energy $E_{F}$ for a fixed incident
angle $\theta=0$, $V_{g}$ = 5 mV (Black line) and $V_{g}$ = -5 mV(Blue line).
$E_{g}$ denotes the bulk gap of HgTe QW. The other parameters used in the
calculation are $A=364.5$meV$\cdot$nm, $B=-686$meV$\cdot$nm$^{2}$, $C=0$,
$D=-512$meV$\cdot$nm$^{2}$ and $M=-10$ meV.}%
\label{fig:model}%
\end{figure}

Electrons incident perpendicularly to the interface can be perfectly
transmitted by the p-n junction, without any backscattering, due to
the unique band dispersion of HgTe QWs. This feature is caused by
the helicity of the band structure, and is very similar to Klein
tunneling in graphene. However, there are a number of important
differences between a TI and graphene\cite{RMP} : (1) the spin
$\sigma$ in the Dirac Hamiltonian of graphene denotes a pseudospin
referring to the sublattices, while in a TI $\sigma$ is a genuine
electron spin; (2) In the TI, interband tunneling is suppressed
because of the different energy dispersions in conduction and
valence band, i.e., different group velocities. In fact, the
carriers will be completely reflected when the incident angle is
larger than the critical angle $\theta_{c}$. The critical angle
$\theta_{c}$ can be determined by Snell's law,
$\sin{\theta_{C}}/\sin{\theta_{R}}=k_{F}^{R}/k_{F}^{L}=n_{R}/n_{L}$,
and thus $\theta_{c}=\arcsin{(k_{F}^{R}/k_{F}^{L})}$. An electron in
the TI thus behaves as a photon which is injected from a material
with larger refractive index $n_{2}$ into a medium with smaller
refractive index $n_{1}$. This critical angle for electrons can be
tuned significantly by changing the gate voltage $V_{g}$ (see the
solid curves in Fig. 1(c)), while the critical angle is difficult to
change for photons. As a next step, we now include the RSOI
generated by a perpendicular electric field under the electric gate
(in the gray region of Fig. 1(a)). It is in principle possible to
tune the potential height and the strength of the RSOI independently
utilizing top- and back- gates.\cite{Nitta} Interestingly, the RSOI
leads to a spin-dependent change of the critical angle
$\theta_{c}^{\sigma}$ (see Fig. 1(c)); this is because the RSOI
induces a spin splitting in the band structure, resulting different
Fermi wave vectors $k_{F}^{\sigma}$ ($\sigma=\uparrow,\downarrow$)
(see the solid curves in Fig. 1(b)) for the two subbands. This
implies that it is possible to induce spin-dependent reflection at
the p-n junction, which allows for a fully spin-polarized tunneling
current when the incident angle $\theta$ of the electrons is tuned
properly (see the inset in Fig. 1(c)), i.e., $\theta
_{\downarrow}>\theta>\theta_{\uparrow}$, where $\theta_{\uparrow}$
($\theta_{\downarrow}$) is the critical angle for spin-up (-down)
electron. The difference between the critical angles $\theta_{c}$
with and without the RSOI can be tuned significantly by adjusting
the gate voltage. This allows the realization of a spin refractive
device, as a the building block for spin-optics. When we tune the
Fermi energy for a fixed RSOI strength and incidence angle $\theta$,
the transmission exhibits a gap in which electron tunneling is
forbidden; the width of the gap $E_{g}^{T}$ corresponds to the total
bulk gap of the inverted HgTe QW (see Figs. 1(b) and 1(d)). Fig.
1(d) shows that an initially opaque medium can suddenly become
transparent by adjusting the Fermi level slightly. This rapid
switching between opacity and transparency is a unique feature of
the HgTe planar p-n junction.

\begin{figure}[h]
\includegraphics[width=0.99\columnwidth,clip]{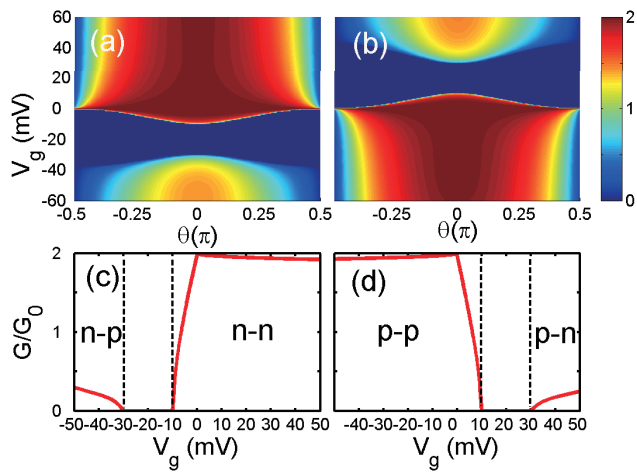} \centering
\caption{(color online). (a) and (b) Contour plot for the electron
transmission as function of incidence angle $\theta$ and gate voltage $V_{g}$
for Fermi energies $E_{F}=\pm20$ meV in the ungated region, respectively. (c)
and (d) Gate voltage $V_{g}-$dependence of conductance $G$ for the Fermi
energy $E_{F}=\pm20$meV, respectively. Here the potential profile of the pn
junction is modeled by a smooth potential.}%
\label{fig:mode2}%
\end{figure}

In order to see how the tunneling behavior can be controlled by adjusting the
incidence angle and the Fermi energy on both sides of the junction, we show
the $\theta$ and $V_{g}$-dependent transmission in Fig. 2. One observes
clearly that perfect transmission occurs at small incidence angles $\theta
_{c}\rightarrow0$ for intraband tunneling process, e.g., n-n and p-p process
(see Figs. 2(a) and 2(b)). In the vicinity of the gap where the tunneling is
forbidden, the critical angle $\theta_{c}$ depends strongly on the gate
voltage $V_{g}$ that tunes the Fermi level on the right hand side of the
junction. The perfect intraband transmission thus could provide us a new
possible experimental method to map the band structure in the vicinity of the
bulk gap. As is evident from Fig. 2(c) and 2 (d), the conductance shows a
strongly asymmetry behavior when comparing intraband and interband tunneling.
For n-n or p-p processes, the perfect transmission leads to a large
conductance while the small interband transmission, i.e., the n-p process,
gives rise to a small conductance. This interesting feature are in good
agreement with a very recent experiment, (Fig. 2.4 in Ref. 22) the
quantitative difference comes from the simplified model used in the
calculation in which the contribution from the light-hole is neglected when
the Fermi energy locates at the bulk states.

\begin{figure}[h]
\centering
\includegraphics[width=0.99\columnwidth,clip]{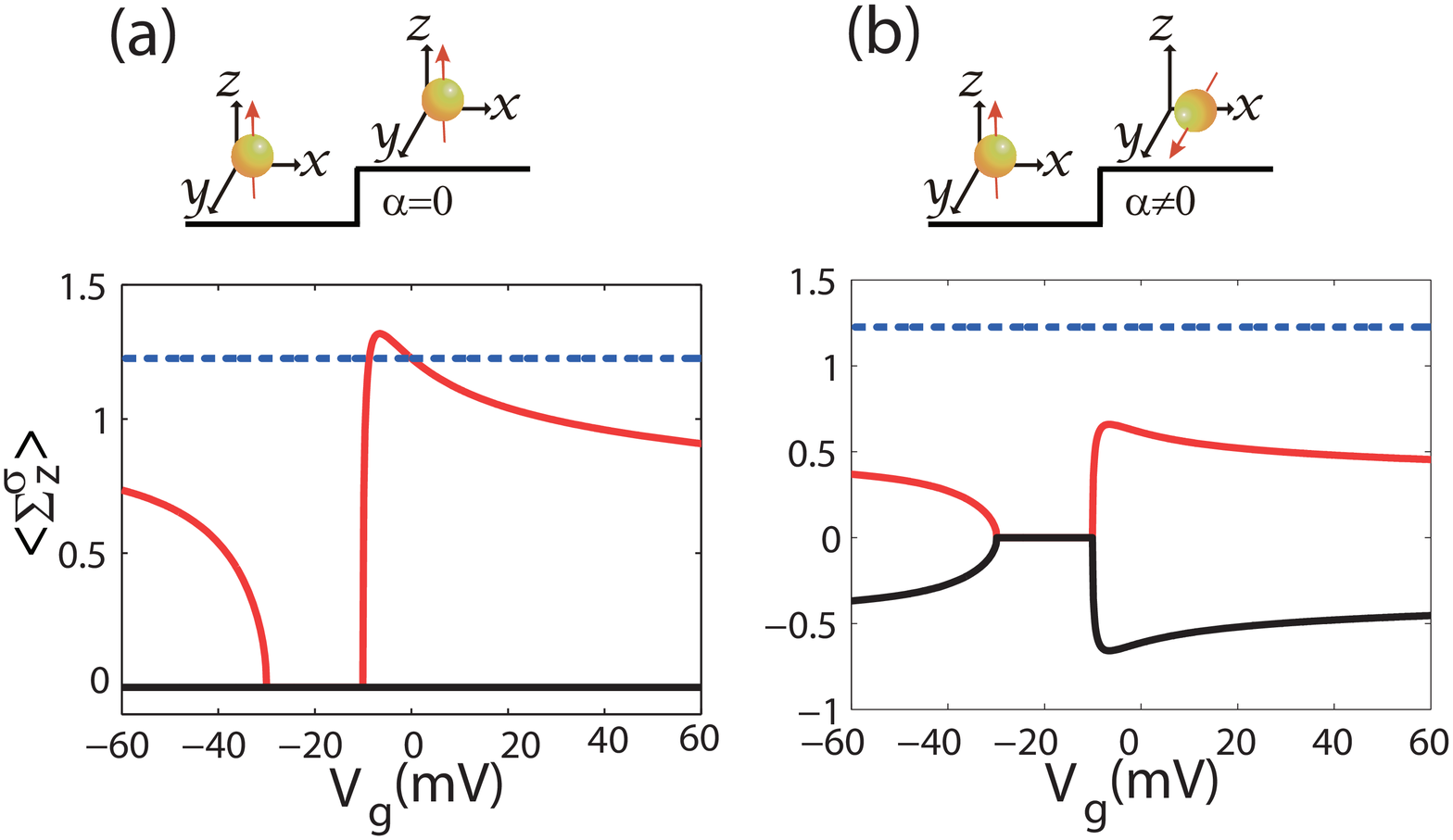}\caption{(color
online). The spin projection $<\Sigma_{z}^{\sigma}>$ as a function of gate
voltage $V_{g}$ (in (a) and (b)) for a fixed Fermi energy $E_{F}=20$ meV and
spin orientation at the left side of p-n junction, which is indicated by the
dashed-blue lines. The red and black lines correspond to the spin projections
$<\Sigma_{z}^{\uparrow}>$ and $<\Sigma_{z}^{\downarrow}>$ of the transmitted
electron, respectively. The insets show schematically the changes of the spin
orientations for the cases without (a) and with (b) RSOI, respectively. The
other parameters are (a) $\theta=0$, $\alpha=0$, (b) $\theta=0$, $\alpha
=5$meV$\cdot$nm, respectively.}%
\label{fig:mode3}%
\end{figure}

We further investigate the spin transport properties of the p-n junction by
calculating the spin projection $<\Sigma_{z}^{\sigma}>=<S_{z,e}^{\sigma
}>+<S_{z,h}^{\sigma}>$, where $<S_{z,e}^{\sigma}>$ ($<S_{z,h}^{\sigma}>$)
denotes the $z$-component of electron (heavy hole) spin. When the RSOI is
present in the system, which, like the Fermi energy, can be tuned by voltage
imposed on the structure (in the right gray region of Fig. 1(a)), the spin
orientation of transmitted carriers changes greatly even for very weak RSOI
($\alpha=5$meV$\cdot$nm). For example, consider an electron injected from the
left side carrying up-spin through the p-n junction. When the RSOI is absent
(see Fig. 3(a)), the gate voltage $V_{g}$ only slightly changes the spin
projection $<\Sigma_{z}^{\sigma}>$. In the presence of a moderate RSOI
($\alpha=5$meV$\cdot$nm), the spin orientation of transmitted electrons
becomes in-plane, since the out-of-plane component $<\Sigma_{z}>=<\Sigma
_{z}^{\uparrow}>+<\Sigma_{z}^{\downarrow}>$ vanishes due to the effective
magnetic field resulting from the RSOI (see Fig. 3(b)). This demonstrates that
the RSOI can substantially rotate the spin orientation.

Up to now, we only have considered two-dimensional case and ignored the
contribution from the helical edge states in the region where the Fermi level
is in the gap. Next, we turn to discuss the quantum tunneling through a p-n
junction in a quasi-one-dimensional (Q1D) channel where the transverse modes
are quantized like in a quantum point contact, but the channel width is still
large enough that the helical edge states localized at opposite edges only
overlap slightly. Such a Q1D QSH bar structure, which could be fabricated
using standard lithographical techniques, is shown schematically in the inset
of Fig. 4(a). Assuming hard-wall confinement, the wave function in the left
side of the structure can be written as $\psi_{L}(x<0)=\chi_{n}^{L}%
e^{ik_{n}^{L}x}\varphi_{n}(y)+\sum_{mn}r_{m}\chi_{m}^{L}e^{-ik_{m}^{L}%
x}\varphi_{n}(y)$, emerging on the right side as $\psi_{R}(x>0)=\sum_{mn}%
t_{m}\chi_{m}^{R}e^{-ik_{m}^{R}x}\varphi_{n}(y)$, where $\varphi_{n}%
(y)=\sqrt{\frac{2}{W}}\sin \frac{n\pi y}{W}$. Using scattering matrix theory
and the Landauer-B$\ddot{u}$ttiker formula we can calculate the total
transmission $T$.

\begin{figure}[h]
\centering
\includegraphics[width=0.99\columnwidth,clip]{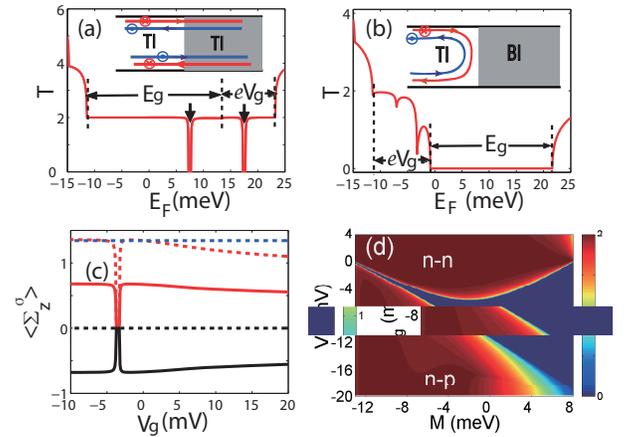}\caption{(color
online). (a) and (b) Fermi energy dependence of the transmission $T$
(red line) in a QSH bar with gate voltage $V_{g}=-10$ mV for TI-TI
and TI-BI tunneling. $E_{g}$ ($eV_{g}$) denotes the gap in the bulk
(the height of the potential profile of the p-n junction), ranging
from -11 meV to 13 meV. The insets show schematically the edge
channels in the TI/TI and TI/BI p-n junctions in the QSH bars. (c)
The spin projections $<\Sigma_{z}^{\sigma}>$ as a function of gate
voltage $V_{g}$ for a fixed Fermi energy $E_{F}=10$ meV. The
dashed-blue lines denote the spin orientation of the incident
electrons. The solid and dashed lines correspond to the cases with
and without RSOI respectively, where the red and black curves
represent the spin $<\Sigma_{z}^{\uparrow}>$ and $<\Sigma
_{z}^{\downarrow}>$. (d) Contour plot of the transmission as
function of $M$ and $V_{g}$ for a fixed Fermi Energy $E_{F}=10$ meV.
The width of the QSH bar
is set at $W=200$ nm. }%
\end{figure}

The helical edge states lead to a conductance plateau at
$G=2e^{2}/h$ when the Fermi energy is located within the bulk gap.
When electrons tunnel through a p-n junction including edge states,
perfect transmission can be observed for the intra- and inter-band
tunneling processes due to the conservation of the helicity of the
edge states. This perfect transmission in a Q1D system differs from
that in 2D discussed above (see Fig. 2). The minigaps in Fig. 4(a)
indicated by the vertical arrows in the plateau are caused by the
coupling between the counter-propagating edge states in the n- and
p-regions with opposite spins stemming from the opposite sides of
the QSH bar. The most interesting aspect revealed by our calculation
is that the conductance plateau due to the edge states is
independent of gate voltage $V_{g}$ due to the conservation of the
helicity of the edge states. A pronounced perfect transmission
feature can also be observed in the other tunneling processes, i.e.,
p-p, n-n, p-n and n-p junctions. In a Q1D, this feature is quite
distinct from that in a 2D system (see Fig. 2). Surprisingly, the
tunneling process between BI and TI, which is shown in Fig. 4(b), is
very different. By tuning the band parameter $M$
electrically,\cite{Kai} the transport through the topological edge
channels can be blocked for the TI/BI ($M>0$) hybridized structure
and become vanishing small for the topological insulator ($M<0$)
(see the insets of Figs. 4(a) and 4(b)). When the electrons are
transported in a TI system ($M<0$), the conductance plateau does not
change with gate voltage $V_{g}$ except for the region of the
minigaps. While for $M>0$, the conductance plateau disappears
totally, because the BI blocks the edge channels. This means that
one can switch on/off the transport property electrically in a
hybrid BI/TI system. Interestingly, the RSOI would not affect the
transmission plateau since the RSOI preserves time reversal symmetry
and would not destroy the edge channels. The spin orientation of the
transmitted electrons can, however, be changed significantly by the
RSOI, just as in the 2D case. In Fig. 4(c), one observes that the
spin projection $<\Sigma_{z}>$ of the transmitted electrons vanishes
for spin-up injection because the RSOI behaves like an in-plane
magnetic field leading to a giant spin rotation.

In summary, we have studied quantum tunneling through p-n junctions in HgTe
QWs with inverted band structures. An interesting perfect transmission of the
quantum tunneling process is found for electrons injected normal to the
interface of the p-n junction. The opacity and transparency of the p-n
junction can be tuned by changing the incidence angle of the incoming
carriers, the gate voltage (which determines the Fermi level on the right hand
side of the junction) and the strength of the Rashba spin-orbit interaction.
Spin-up and spin-down electrons can be separated spatially utilizing the RSOI,
which could result in a building block for spin-optics. Tunneling through
toplogical edge states can be switched on and off by tuning the band
parameters electrically. This provides an efficient means of controlling the
transport properties of topological edge channels electrically.

\begin{acknowledgments}
This work was supported by the NSFC Grant Nos. 60525405 and 10874175. Xie is
supported by US-DOE and US-NSF. K.C. would like to appreciate Prof. R. B. Tao
for helpful discussion.
\end{acknowledgments}

\end{document}